\documentclass{article}
\DeclareUnicodeCharacter{2081}{$_1$}

\usepackage{geometry}
\usepackage{longtable}
\usepackage{amssymb} 
\usepackage{pifont}  
\usepackage{array}
\usepackage{comment}
\usepackage{graphicx}

\usepackage{booktabs}     
\usepackage{siunitx}      

\usepackage{listings}
\usepackage{xcolor}
\definecolor{codegreen}{rgb}{0,0.6,0}
\definecolor{codegray}{rgb}{0.5,0.5,0.5}
\definecolor{codepurple}{rgb}{0.58,0,0.82}
\definecolor{codeblue}{rgb}{0.0, 0.0, 0.8}
\definecolor{codered}{rgb}{0.8, 0, 0}
\definecolor{backcolour}{rgb}{0.97,0.97,0.97}

\usepackage{listings}
\usepackage{xcolor} 

\lstdefinelanguage{json}{
    basicstyle=\ttfamily\small,
    showstringspaces=false,
    breaklines=true,
    literate=
     *{0}{{{\color{blue}0}}}{1}
      {1}{{{\color{blue}1}}}{1}
      {2}{{{\color{blue}2}}}{1}
      {3}{{{\color{blue}3}}}{1}
      {4}{{{\color{blue}4}}}{1}
      {5}{{{\color{blue}5}}}{1}
      {6}{{{\color{blue}6}}}{1}
      {7}{{{\color{blue}7}}}{1}
      {8}{{{\color{blue}8}}}{1}
      {9}{{{\color{blue}9}}}{1}
      {:}{{{\color{black}:}}}{1}
      {,}{{{\color{black},}}}{1}
      {"}{{{\color{red}"}}}{1},
}

\lstdefinestyle{mystyle}{
    backgroundcolor=\color{backcolour},   
    commentstyle=\color{codegreen},
    keywordstyle=\color{codeblue},
    numberstyle=\tiny\color{codegray},
    stringstyle=\color{codered},
    basicstyle=\ttfamily\footnotesize,
    breakatwhitespace=false,         
    breaklines=true,                 
    captionpos=b,                    
    keepspaces=true,                 
    numbers=left,                    
    numbersep=5pt,                  
    showspaces=false,                
    showstringspaces=false,
    showtabs=true,                  
    tabsize=1
}

\newcommand{\cmark}{\checkmark}
\newcommand{\xmark}{\ding{55}}

\usepackage[utf8]{inputenc}
\usepackage[english]{babel}
\usepackage{amsmath}
\usepackage{caption}
\usepackage{subcaption}
\usepackage{float}
\usepackage{amssymb}
\usepackage{amsfonts}
\usepackage{array}
\usepackage{graphicx}
\usepackage{mathrsfs}
\usepackage{multirow}
\usepackage{siunitx}
\newcommand{\bea}{\begin{eqnarray*}}
\newcommand{\eea}{\end{eqnarray*}}
\newcommand{\beao}{\begin{eqnarray}}
\newcommand{\eeao}{\end{eqnarray}}

\usepackage{authblk}
\makeatletter
\renewcommand\AB@affilsepx{, \protect\Affilfont} 
\makeatother
\renewcommand\Affilfont{\small} 

\usepackage{amsmath,amssymb,amsfonts}
\usepackage{amsthm}
\usepackage{xcolor}
\usepackage{nicefrac}
\usepackage{enumitem}
\usepackage{hyperref}
\usepackage{wrapfig}
\usepackage{sidecap}

\usepackage[backend=biber, style=numeric, sorting=none]{biblatex}
\DeclareSymbolFont{symbolsC}{U}{txsyc}{m}{n}
\DeclareMathSymbol{\strictif}{\mathrel}{symbolsC}{74}
\usepackage{xcolor}

\title{Towards a Large Physics Benchmark}
\author[1]{Kristian G. Barman\thanks{Corresponding authors: Sascha Caron (\href{mailto:scaron@nikhef.nl}{scaron@nikhef.nl}) and Kristian G. Barman (\href{mailto:kristiancampbell.gonzalezbarman@ugent.be}{kristiancampbell.gonzalezbarman@ugent.be})}}
\author[2,3]{Sascha Caron$^*$}
\author[2]{Faegheh Hasibi}
\author[2]{Eugene Shalugin}
\author[2]{Yoris Marcet}
\author[2]{Johannes Otte}
\author[2]{Henk W. de Regt}
\author[4,5]{Merijn Moody}

\affil[1]{Ghent University}
\affil[2]{IMAPP and ICIS, Radboud University}
\affil[3]{Nikhef, NL}
\affil[4]{Dutch Institute of Emergent Phenomena, University of Amsterdam}
\affil[5]{Institute of Physics, University of Amsterdam}

\begin{document}

\maketitle
\begin{abstract}
We introduce a benchmark framework developed by and for the scientific community to evaluate, monitor and steer large language model development in fundamental physics. 
Building on philosophical concepts of scientific understanding  and creativity, we develop a scoring system in which each question is scored by an expert for its correctness, difficulty, and surprise. The questions are of three forms: (i) multiple-choice questions for conceptual understanding, (ii) analytical problems requiring mathematical derivation, and (iii) open-ended tasks requiring complex problem solving. Our current dataset contains diverse set of examples, including a machine learning challenge to classify high-energy physics events, such as the four top quark signal.  To ensure continued relevance, we propose a ``living'' benchmark, where physicists contribute questions, for instance alongside new publications. We invite contributions via: 
\url{http://www.physicsbenchmarks.org/}. We hope that this benchmark will enable a targeted AI development that can make a meaningful contribution to fundamental physics research.

\end{abstract}

\newpage
\section{Introduction}

The rapid development of Large Language Models (LLMs) has led to a growing interest in assessing their capabilities in various domains. While there are a variety of benchmarks for general purposes, there is still a lack of benchmarks that assess specific scientific understanding and creativity,
particularly in fundamental physics. Current benchmarks often fall short in several key dimensions when evaluating sophisticated scientific reasoning. First, they typically lack the depth necessary to assess understanding beyond undergraduate or master's level knowledge. Second, they rarely distinguish between mere knowledge retrieval and genuine scientific reasoning. Third, they are often susceptible to gaming \cite{singh2025leaderboard}. Fourth, they seldom incorporate evaluation metrics for the (potential) novelty, surprise, and usefulness of the output (i.e., creativity). Finally, there is still no discussion on how to build a large, community-based and long-lasting benchmark in physics.

This paper introduces a framework for a benchmark designed specifically for the fundamental physics research community to evaluate both the scientific understanding and creative capabilities of LLMs in physics. The benchmark spans three formats (multiple-choice, analytically unique answers, and open-ended code challenges) and includes expert ratings of question difficulty and surprise.

Our goal is to equip both the AI and physics communities with a benchmark that not only measures current capabilities, but also serves as a guidepost for future progress, i.e., toward models that can meaningfully contribute to scientific advancement. 
The idea for such a framework is a joint development of philosophers of science and physicists and is a further development of our discussion of scientific understanding of humans and AI \cite{barman2024towards} and the proposal of large physics models \cite{barman2025large}. Our contribution can be summarized as:

\begin{itemize}
\item Philosophically grounded methodology: We draw from contemporary philosophy of science to  operationalise scientific understanding and creativity. 
\item Benchmark development: We develop a framework for a benchmark developed and evaluated by  experts in fundamental physics, incorporating a human-in-the-loop pipeline for both question creation and evaluation. 
\item Multi-Format evaluation: The benchmark spans multiple question formats to  assess a model’s physics capabilities, from factual recall to complex problem-solving. Model responses are evaluated not only for correctness but also for their difficulty and innovative quality (e.g., the surprise or novelty of the answer), which are taken to be proxies of scientific understanding and creativity. 
\item Question examples: We present examples of different question types, in particular an evaluation of the coding capabilities of LLMs for a real scientific problem in particle physics,
the selection of events with 4 top quarks from the background.

\end{itemize}

The paper is organized as follows: Section 2 reviews existing scientific question-answering benchmarks  and their limitations. Section 3 provides background on philosophical definitions of scientific understanding and creativity, and relates it to the types of questions and scoring methodology. Section 4 describes the practical framework implementation for benchmark generation. Section 5 presents example questions and preliminary results. Section 6 focuses on evaluating models and benchmark generation, including performance reporting for different LLMs.

\section{Benchmarks for Physics}

Several benchmarks have emerged to evaluate LLMs in scientific contexts, each with varying scope and depth. SciQAG \cite{wan2024sciqag} presents a system that automatically generates over 188,000 question-answer pairs from scientific literature across 24 domains, evaluated via criteria like relevance and completeness. While scalable, its focus on factual Question-Answering (QA) and minimal expert validation limits its ability to measure deep understanding. Similarly, GPQA \cite{rein2024gpqa} provides 448 graduate-level, multiple-choice questions in physics, biology, and chemistry, crafted by PhD-level contributors to resist search-based answers. Though rigorous, it remains restricted to multiple-choice  questions, as extensive, non-automated evaluation of questions makes it complicated to scale and update the dataset. SciEval \cite{sun2024scieval} and SciFact \cite{wadden2020fact} are broad multi-domain and claim-verification datasets that primarily test surface-level reasoning and factual consistency rather than generative or creative capabilities.

Recent efforts like BRIGHT and SchNovel move beyond traditional formats. BRIGHT \cite{su2024bright} introduces a reasoning-intensive retrieval benchmark that challenges models to identify relevant documents based on underlying principles and multi-step logic, rather than relying on surface-level lexical or semantic similarity. SchNovel \cite{lin2024evaluating} evaluates a model's capacity to assess novelty in scientific research by comparing paired scholarly papers published years apart, assuming temporal novelty progression, and introducing a RAG-based simulation of peer review for improved prediction accuracy. 

Humanity’s Last Exam (HLE) \cite{phan2025humanity} is one of the most challenging evaluation of LLMs’ scientific reasoning across modalities and disciplines (partly due to question difficulty  and uncertainty-awareness). While HLE focuses on structured academic problems and emphasises correctness under uncertainty, our benchmark is designed to complement this by targeting deeper dimensions of understanding and creativity. Additionally, we introduce a multi-format evaluation that includes open-ended problem solving, counterfactual reasoning, and code-based challenges, enabling assessment of a model's capacity not only to reason accurately, but to generate novel and valuable scientific ideas. Where HLE tests mastery over known content, our approach aims to measure how far models can go in pushing the boundaries of scientific inquiry. 

Among recent efforts, TPBench (Theoretical Physics Benchmark) \cite{chung2025theoretical} stands out for focusing specifically on high-energy theory and cosmology, offering 57 carefully curated problems ranging from undergraduate to research level. While TPBench successfully introduces novel problems and emphasises challenges like auto-verifiability and holistic AI grading, its small dataset size and static problem set inherently limit scalability compared to our approach. Unlike continuously updated framework that systematically collects new, expert-level problems and integrates diverse question types (including code-based challenges), TPBench is constrained by a single release of a modest-sized dataset. Additionally, while TPBench focuses on theoretical physics, it does not explicitly incorporate tasks to evaluate creativity. 

Most existing benchmarks focus on evaluating model accuracy, typically through multiple-choice or short-form QA tasks where success is defined by selecting or producing the correct answer. While useful for measuring surface-level competence, this approach often overlooks deeper dimensions of understanding and creativity. Selecting the right option doesn’t necessarily imply reasoning or insight, and incorrect responses in the open-ended questions offer little granularity in assessing the nature of a model’s mistake. In contrast, our benchmark is not designed to ask ``how accurate is the model on this dataset,'' but rather \textit{``how well does the model understand the subject?''} and \textit{``how creative can it be within that domain?''} Our goal is to evaluate not only the correctness of given answers but also the capacity to generate novel insights, shifting the focus from only correctness under uncertainty to the broader, more human-like capabilities of understanding and scientific creativity. 

We introduce tasks that are carefully constructed and rated by physics experts along the dimensions of difficulty and surprise. Our benchmark introduces three distinct types of questions (multiple choice, analytically unique, and open ended problem-solving challenges) to capture a wider spectrum of cognitive and creative abilities in LLMs. We explain these in Section \ref{sec:qframework}. A summary comparison of existing benchmarks is presented in Table 1, highlighting the distinctive contributions and improvements introduced by our approach.

\begin{table}[ht]
\centering
\footnotesize
\begin{tabular}{|p{4cm}|c|c|c|c|c|c|c|c|c|c|c|c|}
\hline
\textbf{Dataset} & 
\rotatebox{90}{Multi-domain} & 
\rotatebox{90}{Dynamic / Evolving} & 
\rotatebox{90}{MCQ} & 
\rotatebox{90}{Open-ended QA} & 
\rotatebox{90}{Code-based Tasks} & 
\rotatebox{90}{Reasoning-focused} & 
\rotatebox{90}{Expert Authored} & 
\rotatebox{90}{Expert Validated} & 
\rotatebox{90}{LLM Evaluated} & 
\rotatebox{90}{Postgrad Difficulty} & 
\rotatebox{90}{Creativity Assessed} & 
\rotatebox{90}{Philosophical Grounding} \\
\hline
\textbf{SciQAG-24D \cite{wan2024sciqag}} & \cmark & \xmark & \xmark & \cmark & \xmark & \xmark & (P) & (P) & \cmark & \xmark & \xmark & \xmark \\
\hline
\textbf{GPQA \cite{rein2024gpqa}} & \cmark & \xmark & \cmark & \xmark & \xmark & \cmark & \cmark & \cmark & \xmark & \cmark & \xmark & \xmark \\
\hline
\textbf{SciEval \cite{sun2024scieval}} & \cmark & \cmark & \cmark & \cmark & \xmark & \cmark & (P) & \xmark & \cmark & \cmark & Limited & \xmark \\
\hline
\textbf{SciFact \cite{wadden2020fact}} & \cmark & \xmark & \xmark & \cmark & \xmark & \xmark & \cmark & \cmark & \cmark & \cmark & \xmark & \xmark \\
\hline
\textbf{BRIGHT \cite{su2024bright}} & \xmark & \xmark & \xmark & \cmark & \xmark & \cmark & \cmark & \cmark & \cmark & \cmark & \xmark & \xmark \\
\hline
\textbf{SchNovel \cite{lin2024evaluating}} & \xmark & \xmark & \xmark & \cmark & \xmark & \cmark & \cmark & \cmark & \cmark & \xmark & \cmark & \xmark \\
\hline
\textbf{HLE \cite{phan2025humanity}} & \cmark & \xmark & \cmark & \cmark & \xmark & \cmark & \cmark & \cmark & \cmark & \cmark & Limited & \xmark \\
\hline
\textbf{TPBench \cite{chung2025theoretical}} & \xmark & \xmark & \xmark & \cmark & \cmark (auto-verifiable) & \cmark & \cmark & (P) & \cmark & \cmark & Limited & \xmark \\
\hline
\textbf{Ours (This Work)} & \xmark & \cmark & \cmark & \cmark & \cmark & \cmark & \cmark & \cmark & \cmark & \cmark & \cmark & \cmark \\
\hline
\end{tabular}
\caption{Comparison of scientific benchmarks evaluating LLMs across dimensions such as reasoning, creativity, and philosophical grounding. (P) = partial or implicit expert involvement.}
\label{tab:benchmark-comparison}
\end{table}

\section{Measuring Scientific Understanding and Creativity}

\subsection{Scientific understanding}

To benchmark scientific understanding in AI, we must first clarify what ``understanding'' actually entails. While it might seem sufficient to assess factual accuracy or problem-solving ability, such measures fall short of capturing deeper scientific understanding; namely, the ability to explain, generalize, and reason beyond surface-level information. This is where philosophy of science becomes relevant: it offers conceptual tools for distinguishing between knowledge, explanation, and understanding, and helps define what it is we are trying to measure when we talk about a model ``understanding'' physics.

According to an influential account\cite{deregt2017}, scientific understanding of a phenomenon stems from it being explained within an intelligible theoretical framework; one that enables scientists to qualitatively predict the target system's behavior (while respecting empirical adequacy and internal consistency). Several philosophers  \cite{grimm2006,grimm2010} \cite{kuorikoski2015},  \cite{barman2024towards} further argue that understanding depends on counterfactual reasoning: the ability to explore how phenomena would behave under different conditions. Together, these perspectives converge on a definition of understanding not as the passive possession of facts, but as the active capacity to apply, explain, and reason within theoretical systems. In this light, QA would involve not testing what a model knows but how it can use knowledge flexibly and insightfully. Hence, assessing understanding requires questions of a varying degree of difficulty, where difficulty is not just a practical label but reflects the epistemic structure of the task. When domain experts rate a question as more difficult, they are assessing the level of conceptual engagement required. But more importantly, certain tasks require going through key steps, without which, the correct answer cannot be obtained consistently ( for an example see Type 2 questions in Section~\ref{subsec:type2})

These definitions are not new, but this is the first attempt to integrate them into a scalable, operational benchmark for LLMs through a question-based evaluation that encodes \textit{difficulty} as a proxy for epistemic depth in such a way that answering said questions \textit{accurately} reflects the degree of scientific understanding.

\subsection{Scientific Creativity}

In philosophy, creative products are typically defined as novel, valuable, and surprising \cite{ElliotSTokes}. The necessity of the novelty condition is obvious, as it captures core intuitions about creativity. An important distinction \cite{boden2004creative} is between H-novelty (historical novelty), where a product is new in all of history, and P-novelty (psychological novelty), where it is new to the creator’s mind. While P-novelty applies to humans, it cannot be easily extended to LLMs, as they are trained on vast datasets, often containing the entire internet, but cannot explicitly memorize all this data. For a given problem, it is difficult to determine whether an LLM recalls a solution or generates it differently. To circumvent this issue, we extend P-novelty to LLMs by defining an LLM-generated product to be P-novel if it is novel with respect to the LLMs training data. Since P-novelty has anthropomorphic connotations, we adopt Boden's \cite{bodenbio} I-novelty (individual novelty), which generalizes to non-human entities. So we call an LLM's I-novel if it is novel relative to its training data.

The concept of novelty requires elaboration because, when taken in the strictest sense, it is a weak and ill-defined concept: strictly speaking one can change one word in a book and create a novel text, which is not a creative endeavor. In reality almost all created products are H-novel. To exclude cases of trivial novelty, philosophers \cite{boden2004creative} introduced the \textit{surprise} condition. Surprise measures how well the product can be explained using pre-established generative principles and one can view it as a more sophisticated measure of novelty. For instance, a mathematical proof or physical derivation can follow a well-known step plan, which we would not say to be very creative. This step plan is part of the generative principles of the theory, i.e., it is a known set of rules that can be applied in specific situations to create new products within the theory. Producing products in such a way is not creative. What is creative, is going beyond pre-established generative principles, for instance by using a new proof method, a new trick or by applying methods in areas where they were previously not known to be useful. For surprise we can make the same distinction between H-surprise and I-surprise for LLMs, where H-surprise means a product is surprising with respect to all of history and I-surprise means a product is surprising with respect to the LLM's training data.

The value condition is added to rule out surprising nonsense from being creative: anyone can create something that does not follow any generative principles by generating random sequences, but we would not call this creative \cite{gaut2010philosophy, gaut2018value}. Nonetheless, we should also not limit `value' to societal benefit \cite{gaut2010philosophy, gaut2018value,NovitzExp, Novitz1999-NOVCAC}. A general and apt definition of value is therefore the notion of {characteristic value}\cite{mcmullinValuesScience1982} \textit: ``a property
or set of properties may count as a value in an entity of a particular kind because it is desirable for an entity of that kind.'' Since this is contextually determined, in our case we will mostly associate value with how correctly an answer answers a question.

\subsection{A Question-Based Benchmark Framework}
\label{sec:qframework}

Leveraging this philosophical basis of scientific understanding and creativity, we introduce a scalable and (semi-)automated framework for evaluating LLMs using distinct question types. We base our question types loosely on  \cite{barman2024towards}, where the broader philosophical reasoning for these question types is made explicit and argued for. 

The three types of questions generated in this benchmark can be categorized into the following categories: multiple choice, analytically unique, and open-ended coding challenges. Examples of all types of questions, including prompting strategies, can be seen in Section \ref{sec: examples}. Each of these three question types will range from simple problems to real frontier scientific problems.

\textbf{Type 1: Multiple Choice.}
The simplest question type consists of multiple choice answers (A, B, C, D), with exactly one correct and three incorrect answers. This format allows for straightforward, easily scalable, and automated question generation and evaluation.  

\textbf{Type 2: Analytically Unique.} The second type of question is similar to the first except that no answers are given and the answer is limited to a mathematically unique expression. A step-by-step solution should be provided, not just the end result. This format allows the use of open ended questions without introducing complex and intricate nuances of natural language. Concealing the correct answer allows for the elimination of possible biases in next-token prediction inference when the answer is included in the prompt. 

\textbf{Type 3: Open-Ended Coding Challenges.}
The third question type takes the form of open-ended, code-based problem-solving challenges where the primary objective is to maximize a scalar performance metric (score). Type 3 questions prompt the LLM to generate (Python) code that addresses specific physics problems. Each challenge question will be accompanied by detailed yet deliberately limited context, including clear problem and data descriptions, alongside an elaborate code template that carefully balances detailed instructions with the necessary freedom to foster creative code generation.

Type 3 questions challenge the model with the construction and linking of various steps in a (machine learning) code base. This may include data pre-processing, feature engineering, model design, training procedures, process simulation, and parameter fitting. Rather than testing isolated skills, these tasks assess the model’s ability to reason through and construct a structured workflow that mirrors real scientific problem solving. Because the model can, from a one-shot prompt, generate the whole code, it is feasible to test the whole process. More importantly, because its results can be easily quantified, it is easy to show how close, or even if they can surpass human abilities per single question. General rules that this format must follow are:
\begin{itemize}
    \item Single scalar metric: Evaluation results are determined by a clearly defined metric in the form of a single score ([0,1]) with higher scores indicating better performance.
    \item Sand-boxed execution: Model training and inference must be compartmentalized and isolated from evaluation to ensure strict separation of stages. This is to prevent any form of data leakage or cheating.
    \item Single-prompt, zero-shot: Models receive exactly one prompt without iterative refinement, mitigating prompting biases and memorization.
\end{itemize}
Additionally, note that scoring should not be strongly dependent on hardware to ensure a fair comparison between runs. Moreover, models should be granted sufficient freedom to implement their desired solutions within reasonable bounds that will enable testing the solution.  

\subsection{Scoring Methodology}
\label{subsec:scoring}

The evaluation of LLM performance in this benchmark framework translates into quantifiable scores for scientific understanding and creativity, rated on a 1-5 scale. The scoring approach differs slightly for Type 1/2 questions versus Type 3 challenges. The following details the application of our criteria to Type 1 and Type 2 questions.

For \textit{scientific understanding}, the core metric is \textit{difficulty}. To translate this into a score we let domain experts assign a difficulty score of (1-5) to each question. A higher score indicates that more understanding is required to answer the question correctly. While there is some leeway in how evaluators may subjectively assign difficulty, with enough (or representative enough) evaluators, the differences in interpretation will average into what the community believes is 'difficult'. 

For \textit{scientific creativity} we have to evaluate whether the answers are \textit{surprising} and \textit{valuable}. To measure \textit{value}, we have to specify the characteristic value of the answer to a question. The main value of an answer in our context is the \textit{correctness} of the answer. Fortunately, the correctness is precisely what is evaluated in the benchmark. So value here is merely a binary score of $1$ or $0$ that indicates whether the answer was correct or not.

\begin{table}[H]
  \centering
  \begin{tabular}{|c|c|}
      \hline
      Score & Difficulty and Surprise \\
      \hline
      5 & Excellent \\
      4 & High \\
      3 & Good \\
      2 & Reasonable \\
      1 & Minimal \\
      0 & No contribution* \\
      \hline
  \end{tabular}
    \caption{Description of scores to be assigned by experts to individual questions for difficulty and surprise. No contribution is exclusive to Type 3 questions, in order to retain the possibility of nullifying a question's overall value.}
    \label{tab:scores}
\end{table}

Evaluating the surprise of each generated answer directly is unfeasible. As a practical alternative, we let experts evaluate the \textit{known correct answers} of the questions on surprise again with a score of (1-5). In general, surprise should be evaluated as I-surprise, which in this case means surprise with respect to what was available on the internet before the cut-off date of the model. However, if a result is recent or obscure, we also allow questions where the answer might be partly found online; in such cases, this will be noted, and the date from which the answer could be found online will be recorded so that the product can still be counted as I-surprising for models with a cut-off date before the answer appeared online. We then assume that if an LLM is able to solve one of the questions whose answer has been deemed surprising, it has internally recreated at least part of the reasoning of this surprising answer and therefore created a surprising product. This assumption is more reasonable for type 2 questions than for type 1 questions, as it is possible to answer type 1 questions correctly by an educated guess. Note that a similar argument holds for difficulty, i.e., if an LLM can consistently answer type 2 questions that are deemed difficult, it means it can (at least partially) recreate the steps involved. The underlying idea here is that while it's possible to arrive at a correct answer by chance or through flawed reasoning, this becomes increasingly unlikely as the number of examples, the complexity of the reasoning, or the number of steps increases.

For Type 3 questions the situation is slightly different as the solution achieves a scalar score based on a predefined metric. This score itself serves as the measure of its value. Both difficulty and surprise are then derived from this score using a step-function that maps the continuous score to a discrete score of (0-5). For difficulty, experts will define six distinct regions within the possible score range. A rating from 0 to 5 is assigned based on the region into which the model's final score falls, as achieving a higher score on these problems is more difficult. Similarly, for surprise, the score range is partitioned into six regions, each corresponding to a surprise value from 0 to 5. This rating can, for instance, be determined by comparing the model's score against established benchmarks, such as the performance of other models or the state-of-the-art achieved by human researchers. A score that surpasses these known ceilings would thus be assigned a higher surprise value.

\section{Implementation of the Framework}

\begin{figure}[htp]
    \centering
    \includegraphics[width=15cm]{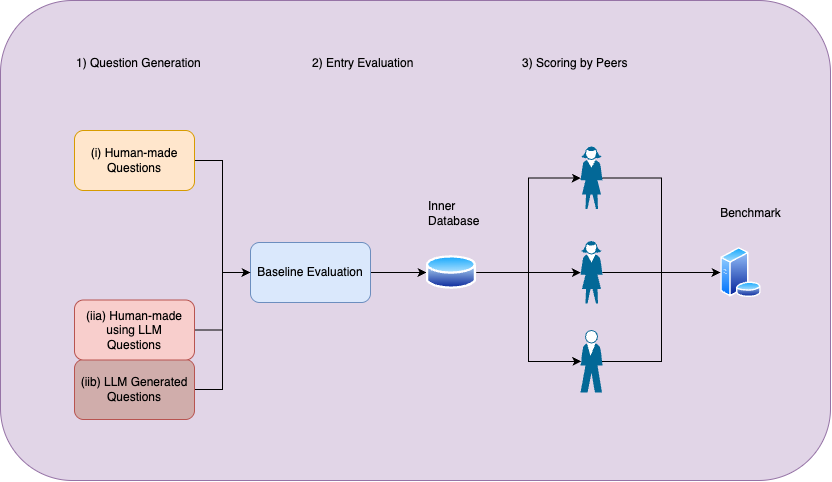}
    \caption{Framework for Physics Creativity and Reasoning Benchmark}
    \label{fig:pipeline}
\end{figure}

To probe a model’s scientific understanding and creativity at scale, we treat question–answer (QA) creation and QA evaluation as distinct tasks. Each QA pair\footnote{For type-3 questions QA-pair generation refers to the entire challenge including sub-questions, but we will call it QA-generation for simplicity.} is scored along two dimensions. The full QA creation and evaluation pipeline follows a three step pipeline, illustrated in Figure \ref{fig:pipeline}.

\textbf{1) Question Generation.} We distinguish between two QA-generation methods. Either questions are generated by (i) exclusively human experts, or by (ii) human experts collaborating with large language models. In the case of the latter, we make an additional distinction between: (ii)a mainly human-generated and LLM assisted, or (ii)b mainly LLM generated and human-assisted. 

\begin{figure}[htp]
    \centering
    \begin{subfigure}[t]{0.39\textwidth}
        \centering
        \includegraphics[height=3.5cm]{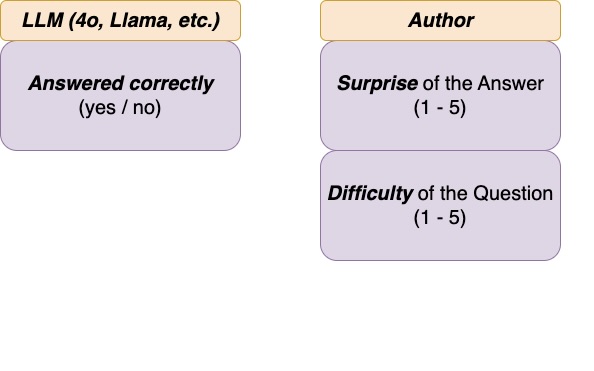}
        \caption{Question Generation Step}
        \label{fig:pipeline_a}
    \end{subfigure}
    \hfill
    \begin{subfigure}[t]{0.59\textwidth}
        \centering
        \includegraphics[height=3.5cm]{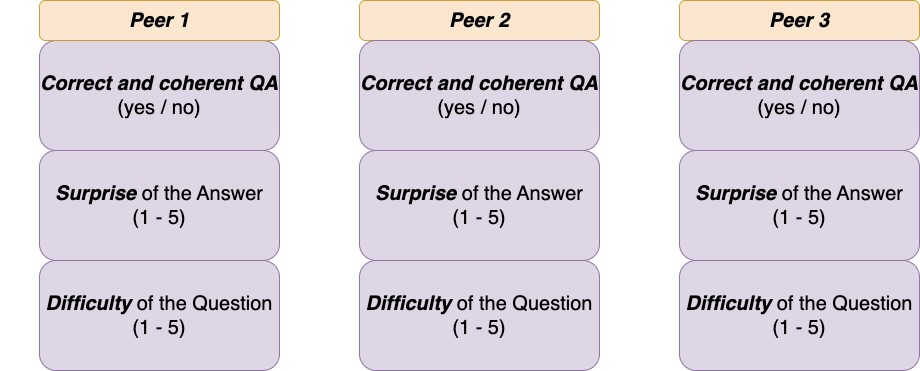}
        \caption{Scoring by Peers Step}
        \label{fig:pipeline_b}
    \end{subfigure}
    \caption{Framework for Physics Creativity and Reasoning Benchmark: (a) Question Generation Step (b) Scoring by Peers Step}
    \label{fig:pipeline_23}
\end{figure}

\textbf{2) Expert Evaluation.} Alongside QA-generation each human expert is expected to score their question on \textit{difficulty} and the correct answer on \textit{surprise}. Additionally, questions will be tested against chosen LLMs as shown in Figure \ref{fig:pipeline_a}. The benchmark will provide information on the ability of chosen models to answer submitted questions. 

\textbf{3) Peer Evaluation.} Each QA-pair will be evaluated by three additional experts on both formal correctness as well as surprise and difficulty as shown in Figure \ref{fig:pipeline_b}. Then, for each metric a final average score will be determined for each QA-pair. The result will be an extensive database of quality questions created and evaluated by the physics community itself. 

Every QA pair will be peer reviewed according to the following points:
\begin{itemize}
    \item \textbf{Author scores} - the author assigns an initial \textit{difficulty} label and rates the correct answer's \textit{surprise} and \textit{difficulty} as defined in section \ref{subsec:scoring}.
    \item \textbf{Triple check} - Each QA pair is verified by, at least, three independent experts on (i) Formal correctness and (ii) \textit{difficulty} and \textit{surprise}. Once a QA has been vetted an average score will be calculated for each of the three figures of merit.
\end{itemize}

If any reviewer flags a fatal flaw, the QA pair is returned for revision or discarded. Overall difficulty distributions are monitored to keep the benchmark balanced. 

\textbf{Benchmark Entry}
After acceptance each QA record stores: the question-answer pair, question type specific information and the different scores. An example of a type 2 question being store in the database can be found in the appendix \ref{appendix:type2json}.

\textbf{Expert Pool}
To maximise the quality of questions, we invite contributors holding a PhD (or equivalent research experience) in fundamental physics (e.g. particle physics, astrophysics, nuclear physics) and supply a detailed question-design manual. \footnote{The manual and a submission portal will be hosted at \url{https://physicsbenchmarks.org}} Contributors will be rewarded for their contribution with co-authorship in future publications.

\textbf{Transparency}
All public questions, scores, and version histories will be published through a versioned release approach and searchable at the project site. This transparency lets the community audit both human and LLM contributions and facilitates re-sampling for future benchmark releases. To maintain the integrity of the benchmark, only a small subset of the question database can be made public with each iteration. Models consisting of tools which have access to the internet must be prohibited to gain access to the QA-pairs, and the questions themselves should not be used for training.

\section{Prompting \& Question Examples} \label{sec: examples}


This chapter provides examples for each of the three question types introduced in section \ref{sec:qframework}. For each type, we present representative physics questions reflecting the distinct format and level of complexity involved. Detailed prompting strategies and templates associated with each question type are documented separately in the appendix ~\ref{sec:appendix}.

\subsection{Type 1 question - multiple choice}
\textbf{Example question 1:}
Why does the Higgs boson decay dominantly to \(b\) quarks?

\textbf{Multiple choice answers:}
\begin{enumerate}
    \item[A.] \(b\) quarks are the lightest quarks.
    \item[B.] The top quark is too heavy for the Higgs decay.
    \item[C.] The \(b\) quarks have the right electric charge.
    \item[D.] The Higgs dominantly decays to photons.
\end{enumerate}

\textbf{Correct answer: B}

\textbf{Example question 2:}
Why did we introduce the Higgs mechanism?

\textbf{Multiple choice answers:}
\begin{enumerate}
    \item[A.] We wanted to give masses to all particles. The Higgs field makes all particles heavier.
    \item[B.] The Higgs field gives mass to the proton and neutron.
    \item[C.] When adding simple mass terms to the theory one encounters mathematical problems (e.g. divergences), and this is solved with the Higgs field.
    \item[D.] There was a problem with helicity. The Higgs field explains why we have only left-handed particles.
\end{enumerate}

\textbf{Correct answer: C}

\subsection{Type 2 question: Open ended with a determined analytical answer}
\label{subsec:type2}


\textbf{Example question 1:}
In a two-body scattering event, \( A + B \rightarrow C + D \), it is convenient to introduce the Mandelstam variables:

\begin{align*}
    s &\equiv (p_A + p_B)^2, \\
    t &\equiv (p_A - p_C)^2, \\
    u &\equiv (p_A - p_D)^2.
\end{align*}

Find the energy of A in the center-of-mass frame of A and B, in terms of s, t, u and the rest masses.

\textbf{Correct answer:}
\[
E_A = \frac{s + m_A^2 - m_B^2}{2 \sqrt{s}}.
\]

\textbf{Example question 2:}
The coupling of the Standard-Model Higgs boson to fermions is described by a vertex factor 
\( im_f / v \) 
where \( m_f \) is the rest mass of the fermion and \( v \) is the vacuum expectation value of the Higgs field 
\((= 2m_W / g_W)\).

Calculate the matrix element \( M \) for the Higgs boson decaying into a fermion/antifermion pair. Express the amplitude as a function of \( m_H \) and \( m_f \), where \( m_H \) is the Higgs mass, and show the average over all possible spin configurations as a final answer (if needed, neglect the color factors).

\textbf{Correct answer:}
\[
\langle |M|^2 \rangle = \frac{2 m_f^2}{v^2} \left( m_H^2 - 4m_f^2 \right).
\]

Preliminary results from evaluating four state-of-the-art large language models (LLMs)— ChatGPT 4o, Gemini 2.5 Pro, DeepSeek v3, and Claude Sonnet 4 — on two type 1 and 2 questions are shown in Table~\ref{tab:llm_eval}.

\begin{table}[h!]
\centering
\caption{Preliminary evaluation of large language models (LLMs) on two types of physics questions. All models answered both multiple choice (Type 1) and analytical (Type 2) questions correctly.}
\label{tab:llm_eval}
\begin{tabular}{|l|c|c|c|c|}
\hline
\textbf{Model} & \textbf{Type 1 Q1} & \textbf{Type 1 Q2} & \textbf{Type 2 Q1} & \textbf{Type 2 Q2} \\
\hline
ChatGPT 4o & Correct (B) & Correct (C) & Correct & Correct \\
\hline
Gemini 2.5 Pro & Correct (B) & Correct (C) & Correct & Correct \\
\hline
DeepSeek v3 & Correct (B) & Correct (C) & Correct & Correct \\
\hline
Claude Sonnet 4 & Correct (B) & Correct (C) & Correct & Correct \\
\hline
\end{tabular}
\end{table}

\subsection{Type 3 Question - Maximizing Score}

The first Type 3 question consists of a challenge of binary classification of signal versus background data of proton-proton collision events. The signal consists of the detection of two top and anti-top quark pairs. An example of a specific challenge question is as follows:

"Write Python code for a binary classification model focusing on maximizing the AUC using the code template above. You may freely choose any pre-processing methods and techniques as well as model architecture and training conventions. Do absolutely everything in your power to achieve the highest possible AUC."

The full prompt, code template and LLM generated code can be found in the appendix, section \ref{appendix:fourtops}.

\subsubsection{Preliminary Results:}

\begin{table}[H]
  \centering
  \begin{subtable}[t]{0.46\textwidth}
    \centering
    \caption{LLM results (23-06)}
    \label{tab:fourtop_results}
    \begin{tabular}{@{}lS[table-format=1.4]@{}}
      \toprule
      \textbf{LLM} & {\textbf{AUC}} \\
      \midrule
      ChatGPT 4o-mini-high  & 0.8175 \\
      ChatGPT o3 Pro        & 0.8221 \\
      Claude Sonnet 4.0     & 0.8179 \\
      Gemini 2.5 Pro        & 0.8469 \\
      X-AI Grok             & 0.8183 \\
      Deepseek Chat v3      & 0.8224 \\
      \bottomrule
    \end{tabular}
  \end{subtable}\hfill
  \begin{subtable}[t]{0.46\textwidth}
    \centering
    \caption{Specialized physics models}
    \label{tab:fourtops_original_results}
\begin{tabular}{@{}lS[separate-uncertainty,uncertainty-mode = compact,retain-zero-uncertainty = true,table-format=1.4(1)]@{}}
      \toprule
      \textbf{Model Type} & {\textbf{AUC}} \\
      \midrule
      PN                            & 0.8471(1) \\
      PN\textsubscript{int.SM}      & 0.8725(0) \\
      ParT                          & 0.8404(0) \\
      ParT\textsubscript{int.SM}    & 0.8732(0) \\
      \bottomrule
    \end{tabular}
  \end{subtable}
  \caption{Side-by-side comparison of preliminary results of LLM performance on the \textit{fourtops} challenge (left) with specialized models (right) on the same \textit{fourtops} dataset. The two specialized models depicted are Particle Net (PN) a graph neural network and Particle Transformer (ParT). Each with and without integrated physics. In case of the former the models include pairwise features (int.) and SM running coupling constant (SM). Full description can be found in reference \cite{fourtops2022pre}.}
\end{table}

\begin{figure}[H]
  \centering
  \includegraphics[width=1\textwidth]{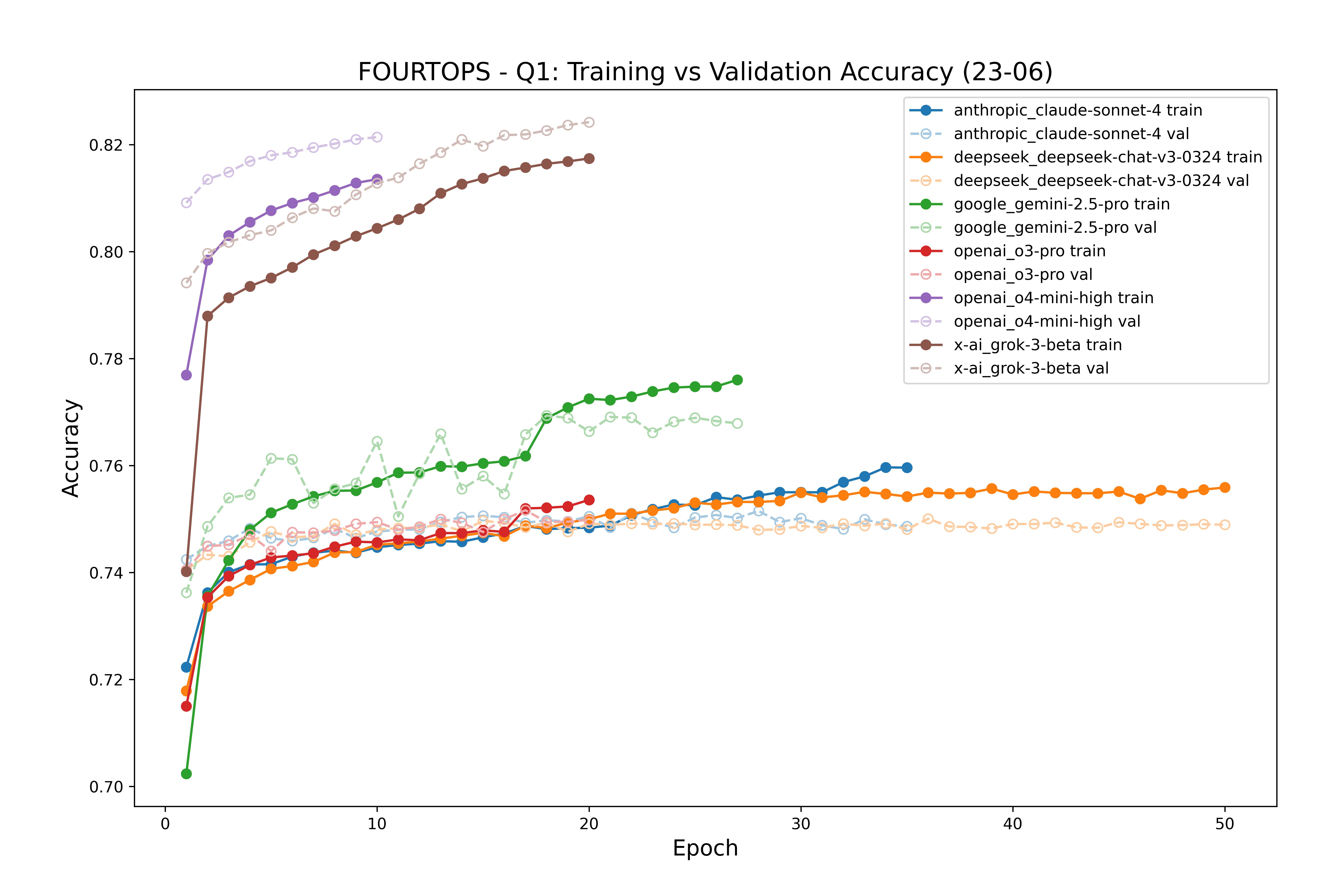}
  \caption{Training and validation accuracy versus epoch for six evaluated models of the example question of the FOURTOPS challenge. The number of epochs changes per model as per their own choice.}
  \label{fig:fourtoptraining}
  \includegraphics[width=1\textwidth]{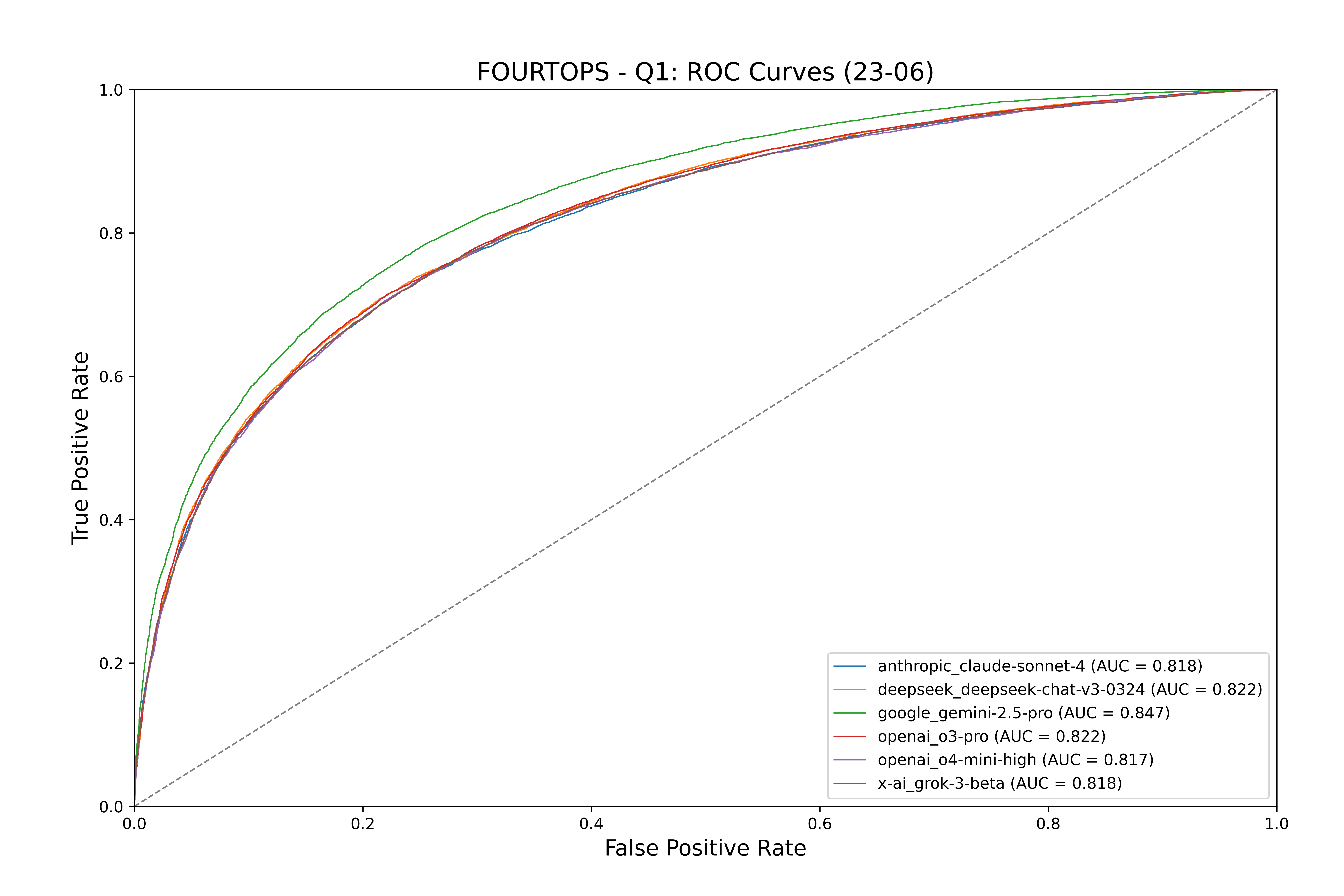}
  \caption{ROC Curves and AUC scores for the example question of the FOURTOPS challenge.}
  \label{fig:fourtoproc}
\end{figure}

\section{Model Evaluation}
\label{sec:eval}

Large language models are evaluated for their scientific understanding and creativity, where scores for \textit{difficulty} and \textit{surprise} serve as proxies for these two metrics, respectively. Each of the three question types produces a separate averaged score for difficulty and surprise, resulting in six total scores. Subsequently, these scores are aggregated into two final values. LLMs will be tested out-of-the-box, meaning we do not consider modifying various parameters such as temperature, nucleus- or top k-sampling. 

Type 1 questions are evaluated through a straightforward comparison between the LLM-generated answer and the known correct answer, yielding a boolean result (correct or incorrect). Type 2 questions, which produce open-ended mathematical expressions, are evaluated using suitable verification software (e.g., Mathematica or numerical Python routines). A suitable verification method must be explicitly specified by the creator of each question. 

Mathematically, the scores for Type 1 and 2 questions can be expressed as follows:

\begin{equation}
    D_{1,2} = \alpha_{1,2} \sum_{i=1}^{N,M} c_i d_i, \quad
    S_{1,2} = \beta_{1,2} \sum_{i=1}^{N,M} c_i s_i, \quad
    c_i =
        \begin{cases}
            1,& \texttt{if correct}\\
            0,& \texttt{if incorrect}
        \end{cases}
\end{equation}

Here, \textit{D} represents difficulty, \textit{S} represents surprise and \textit{c} represents correctness as a proxy for value. The terms $d_i$ and $s_i$ denote the average difficulty and surprise ratings for each individual question, respectively. The normalization weights $\alpha_{1,2}$ and $\beta_{1,2}$ are inversely proportional to the total number of questions per type in the benchmark, under the assumption that the effort of creating questions correlates with their relative epistemic depth. \textit{N} and \textit{M} are the total number of Type 1 and Type 2 questions, respectively. 

For Type 3 questions, evaluation depends directly on a continuous scoring metric specific to each challenge. Let the scoring metric be $c \in (0,1)$. Then the scores for Type 3 questions can be defined as follows:

\begin{equation}
    D_{3} = \alpha_3    \sum_{i=1}^K    d_i, \quad
    S_{3} = \beta_3     \sum_{i=1}^K    s_i, \quad
    d_i,s_i(c) =
        \begin{cases}
        0,       & t^{d,s}_{0}  \le c \le   t^{d,s}_{1} \\[4pt]
        1,       & t^{d,s}_{1}  <   c \le   t^{d,s}_{2} \\[2pt]
        \;\vdots & \\[2pt]
        5,       & t^{d,s}_{4}  <   c \le     t^{d,s}_{5}
        \end{cases}
\end{equation}

Here, $d_i(c)$ and $s_i(c)$ represent disjoint discretized mappings of difficulty and surprise derived from the continuous score $c$, using predefined interval breakpoints $t_{0}<t_{1}<\dots<t_{5} \in (0,1)$. The mapping for surprise is challenge-specific and the mapping for difficulty is question-specific. The difficulty between two questions in the same challenge may differ due to additional contextual information provided or omitted the model. It is important to emphasize that correctness in Type 3 questions is implicitly encoded: even scores greater than zero may reflect no actual contribution. For instance, in a binary classification task, an accuracy score of 0.5 merely indicates random guessing and thus carries no meaningful contribution.

Finally, the scores for each of the three question types can be aggregated into two averaged scores for understanding and creativity:

\begin{equation}
    D_F = \frac{1}{3} \big( D_1 + D_2 + D_3 \big), \quad
    S_F = \frac{1}{3} \big( S_1 + S_2 + S_3 \big)
\end{equation}

The quantities $D_F$ and $S_F$ represent the final scores of the Large Physics Benchmark. Note that it is possible to add further weights to also reflect structural differences in type 1,2,3. At this stage we opt for starting with equal weights, leaving room to gather evidence before committing to a more nuanced scheme.  

\section{Conclusion}

In this paper, we introduced a  benchmark framework designed to assess scientific understanding and creativity in Large Language Models (LLMs) within the domain of fundamental physics.  The benchmark comprises three distinct question types: multiple-choice, analytically unique, and coding challenge tasks that require maximizing a defined metric. Each question is scored along dimensions of difficulty and surprise by domain experts. Our framework incorporates expert-authored and LLM-assisted question creation, human-in-the-loop evaluation, and transparent public documentation, aiming to serve both as a diagnostic tool and a guidepost for future LLM development in scientific reasoning. 

Nonetheless, there are some noteworthy limitations. Identifying duplicate or near-identical problems across large datasets remains a challenge for fair surprise evaluation. Additionally, comparing human and model performance is also difficult, as meaningful comparisons require questions with sufficient context, where both must reason from the same information; replicating such environments is challenging. 

Future work should focus on expanding the diversity and number of high-quality questions, particularly in underrepresented physics subdomains. We aim to improve automatic surprise estimation and explore richer formats for evaluating creativity, such as multi-step reasoning chains or model-generated questions. Incorporating human–model collaboration tasks and longitudinal evaluation of models as they evolve across versions may offer deeper insights into scientific understanding. Additional future directions include refining the benchmark’s ability to distinguish between memorization and reasoning, integrating multi-modal tasks involving visual or symbolic physics representations, and developing adaptive benchmarking tools that adjust question difficulty based on model responses and are ammenable to tayloring context provided. Cross-domain transfer tasks and meta-reasoning challenges may also help evaluate generalization beyond physics.
We invite the physics community to participate in the creation of questions and answers and to become an co-author of a follow-up publication. If you are interested, please visit
\url{http://www.physicsbenchmarks.org/}.

\printbibliography

@inproceedings{sun2024scieval,
  title={Scieval: A multi-level large language model evaluation benchmark for scientific research},
  author={Sun, Liangtai and Han, Yang and Zhao, Zihan and Ma, Da and Shen, Zhennan and Chen, Baocai and Chen, Lu and Yu, Kai},
  booktitle={Proceedings of the AAAI Conference on Artificial Intelligence},
  volume={38},
  number={17},
  pages={19053--19061},
  year={2024}
}

@article{wan2024sciqag,
  author = "Wan, Yuwei and Liu, Yixuan and Ajith, Aswathy and Grazian, Clara and Hoex, Bram and Zhang, Wenjie and Kit, Chunyu and Xie, Tong and Foster, Ian",
  title = "{SciQAG: A Framework for Auto-Generated Science Question Answering Dataset with Fine-grained Evaluation}",
  eprint = "2405.09939",
  archivePrefix = "arXiv",
  primaryClass = "cs.CL",
  year = "2024"
}

@inproceedings{rein2024gpqa,
  author = "Rein, David and Hou, Betty Li and Stickland, Asa Cooper and Petty, Jackson and Pang, Richard Yuanzhe and Dirani, Julien and Michael, Julian and Bowman, Samuel R.",
  title = "{GPQA: A Graduate-Level Google-Proof Q\&A Benchmark}",
  booktitle = "Proceedings of the First Conference on Language Modeling",
  year = "2024",
  eprint = "2311.12022",
  archivePrefix = "arXiv",
  primaryClass = "cs.AI"
}

@article{singh2025leaderboard,
  author = "Singh, Shivalika and Nan, Yiyang and Wang, Alex and D'Souza, Daniel and Kapoor, Sayash and Üstün, Ahmet and Koyejo, Sanmi and Deng, Yuntian and Longpre, Shayne and Smith, Noah A. and Ermis, Beyza and Fadaee, Marzieh and Hooker, Sara",
  title = "{The Leaderboard Illusion}",
  eprint = "2504.20879",
  archivePrefix = "arXiv",
  primaryClass = "cs.LG",
  year = "2025"
}

@article{barman2024towards,
    author = "Barman, Kristian G. and Caron, Sascha and Claassen, Tom and de Regt, Henk",
    title = "{Towards a Benchmark for Scientific Understanding in Humans and Machines}",
    eprint = "2304.10327",
    archivePrefix = "arXiv",
    primaryClass = "cs.AI",
    doi = "10.1007/s11023-024-09657-1",
    journal = "Minds and Machines",
    volume = "34",
    number = "1",
    pages = "6",
    year = "2024"
}

@book{deregt2017,
  author = {de Regt, Henk W.},
  title = {Understanding Scientific Understanding},
  publisher = {Oxford University Press},
  year = {2017}
}

@article{grimm2006,
  author = {Grimm, Stephen R.},
  title = {Is Understanding a Species of Knowledge?},
  journal = {The British Journal for the Philosophy of Science},
  volume = {57},
  number = {3},
  pages = {515--535},
  year = {2006}
}

@incollection{grimm2010,
  author = {Grimm, Stephen R.},
  title = {Understanding},
  booktitle = {The Routledge Companion to Epistemology},
  editor = {Bernecker, Sven and Pritchard, Duncan},
  publisher = {Routledge},
  pages = {84--94},
  year = {2010}
}

@article{kuorikoski2015,
  author = {Kuorikoski, Jaakko and Ylikoski, Petri},
  title = {External Representations and Scientific Understanding},
  journal = {Synthese},
  volume = {192},
  pages = {3817--3837},
  year = {2015}
}

@article{barman2025large,
    author = "Barman, Kristian G. and others",
    title = "{Large Physics Models: Towards a collaborative approach with Large Language Models and Foundation Models}",
    eprint = "2501.05382",
    archivePrefix = "arXiv",
    primaryClass = "physics.data-an",
    month = "1",
    year = "2025"
}

@incollection{ElliotSTokes,
  title={Attributing creativity},
  author={Elliot Samuel Paul, Dustin Stokes},
    year = {2018},
    month = {02},
    pages = {193-209},
  booktitle={Creativity and philosophy},
  editor = {Gaut and Kieran},
isbn = {9781351199797},
doi = {10.4324/9781351199797-6}
}

@incollection{bodenbio,
author = {Margaret Boden},
year = {2018},
month = {02},
pages = {186-192},
title = {Creativity and biology},
  booktitle={Creativity and philosophy},
  editor = {Gaut and Kieran},
isbn = {9781351199797},
doi = {10.4324/9781351199797-6}
}

@article{wadden2020fact,
  author = "Wadden, David and Lin, Shanchuan and Lo, Kyle and Wang, Lucy Lu and van Zuylen, Madeleine and Cohan, Arman and Hajishirzi, Hannaneh",
  title = "{Fact or Fiction: Verifying Scientific Claims}",
  eprint = "2004.14974",
  archivePrefix = "arXiv",
  primaryClass = "cs.CL",
  year = "2020"
}

@article{su2024bright,
  author = "Su, Hongjin and Yen, Howard and Xia, Mengzhou and Shi, Weijia and Muennighoff, Niklas and Wang, Han-yu and Liu, Haisu and Shi, Quan and Siegel, Zachary S. and Tang, Michael and others",
  title = "{BRIGHT: A Realistic and Challenging Benchmark for Reasoning-Intensive Retrieval}",
  eprint = "2407.12883",
  archivePrefix = "arXiv",
  primaryClass = "cs.IR",
  year = "2024"
}

@article{lin2024evaluating,
  author = "Lin, Ethan and Peng, Zhiyuan and Fang, Yi",
  title = "{Evaluating and Enhancing Large Language Models for Novelty Assessment in Scholarly Publications}",
  eprint = "2409.16605",
  archivePrefix = "arXiv",
  primaryClass = "cs.CL",
  year = "2024"
}

@article{phan2025humanity,
  author = "Phan, Long and Gatti, Alice and Han, Ziwen and Li, Nathaniel and Hu, Josephina and Zhang, Hugh and Zhang, Chen Bo Calvin and Shaaban, Mohamed and Ling, John and Shi, Sean and others",
  title = "{Humanity's Last Exam}",
  eprint = "2501.14249",
  archivePrefix = "arXiv",
  primaryClass = "cs.AI",
  year = "2025"
}

@article{gaut2010philosophy,
  title={The philosophy of creativity},
  author={Gaut, Berys},
  journal={Philosophy Compass},
  volume={5},
  number={12},
  pages={1034--1046},
  year={2010},
  publisher={Wiley Online Library}
}

@article{Novitz1999-NOVCAC,
	author = {David Novitz},
	doi = {10.1080/00048409912348811},
	journal = {Australasian Journal of Philosophy},
	number = {1},
	pages = {67--82},
	publisher = {Taylor \& Francis},
	title = {Creativity and Constraint},
	volume = {77},
	year = {1999}
}

@incollection{NovitzExp,
  author    = "David Novitz",
  title     = "Explanations of Creativity",
  booktitle = "The Creation of Art: New Essays in Philosophical Aesthetics",
  publisher = "Cambridge UP",
  year      = "2003",
  editor    = "Berys Gaut and Paisley Livingston",
  pages     = "174-91",
  address   = "Cambridge",
}

@incollection{gaut2018value,
  title={The value of creativity},
  author={Gaut, Berys},
    year = {2018},
    month = {02},
    pages = {124-140},
  booktitle={Creativity and philosophy},
  editor = {Gaut and Kieran},
isbn = {9781351199797},
doi = {10.4324/9781351199797-6}
}

@article{mcmullinValuesScience1982,
  title = {Values in {{Science}}},
  author = {McMullin, Ernan},
  year = {1982},
  journal = {PSA: Proceedings of the Biennial Meeting of the Philosophy of Science Association},
  volume = {1982},
  eprint = {192409},
  eprinttype = {jstor},
  pages = {3--28},
  publisher = {[University of Chicago Press, Springer, Philosophy of Science Association]},
  issn = {0270-8647},
  urldate = {2024-12-22}
}

@book{boden2004creative,
  title={The creative mind: Myths and mechanisms},
  author={Boden, Margaret A},
  year={2004},
  publisher={Routledge}
}

@article{chung2025theoretical,
  author = "Chung, Daniel J. H. and Gao, Zhiqi and Kvasiuk, Yurii and Li, Tianyi and Münchmeyer, Moritz and Rudolph, Maja and Sala, Frederic and Tadepalli, Sai Chaitanya",
  title = "{Theoretical Physics Benchmark (TPBench)—A Dataset and Study of AI Reasoning Capabilities in Theoretical Physics}",
  eprint = "2502.15815",
  archivePrefix = "arXiv",
  primaryClass = "physics.comp-ph",
  year = "2025"
}

@ARTICLE{fourtops2022pre,
  title         = "Attention to the strengths of physical interactions:
                   Transformer and graph-based event classification for
                   particle physics experiments",
  author        = "Builtjes, Luc and Caron, Sascha and Moskvitina, Polina and
                   Nellist, Clara and de Austri, Roberto Ruiz and Verheyen, Rob
                   and Zhang, Zhongyi",
  month         =  nov,
  year          =  2022,
  copyright     = "http://creativecommons.org/licenses/by/4.0/",
  archivePrefix = "arXiv",
  primaryClass  = "hep-ph",
  eprint        = "2211.05143"
}

\section{Acknowledgments}
The work of Caron, De Regt, Barman was supported by an IRP grant from FNWI, Radboud University. Barman's work was also supported by FWO grant 1229124N.

\section{Appendix}
\label{sec:appendix}

\subsection{Type 2 Database}
\label{appendix:type2json}
\begin{lstlisting}[language=json,firstnumber=1]
{
    "author": "<author id>",
    "question": "<text of the question>",
    "answer": "<correct answer>",
    "scoring": {
        "referee_1": {
            "author": "<author id>",
            "correctness": "<score>",
            "difficulty": "<score>",
            "surprise": "<score>"
        },
        "referee_2": {},
        "referee_3": {},
        "baseline_models": {
            "o4": "<solved / not solved>",
            "DeepSeek": "<solved / not solved>",
            "Llama": "<solved / not solved>",
            "..." : "..."
        }
    }
}
\end{lstlisting}

\subsection{FOURTOPS Challenge}
\label{appendix:fourtops}

\subsubsection{Full prompt}
** Instructions **
You are an expert at programming in Python, machine learning,
particle and high energy physics. You will help me answer a question in a machine learning challenge format where you strive to maximise a scalar metric in order to learn more about your scientific creativity and scientific understanding. You will follow all of the instructions 
to your best capabilities. Your first priority is to produce a correct solution in terms of runnable python code. Your second priority is to maximise the scoring metric defined below.

** Problem Description **
A major task in particle physics is the measurement of rare signal 
processes with very small cross-sections. With the unprecedented amount of 
data provided by the upcoming runs of the Large Hadron Collider (LHC), 
one can start to measure these processes. An example is the recent 
observation of four top quarks originating from a single proton-proton 
collision event. Accurate classification of these events is crucial, 
as even a small reduction in background noise on the order of a few tens 
of percent while maintaining the same signal detection efficiency can lead 
to a profound increase in sensitivity.

** Evaluation Metric **
The evaluation metric for this classification task is the area under the curve (AUC),
specified by the area under the receiver operating characteristic (ROC) curve. 
The AUC summarizes a model's ability to distinguish between positive and 
negative classes. The higher the score the better.

** Dataset Description **
The dataset used for this problem consists of simulated proton-proton 
collision at a center of mass energy of 13 TeV. The signal process is defined as 
$pp \rightarrow t \bar{t} t \bar{t}$. The relevant production processes of the
backgrounds are $t \bar{t} + X$ where $X = Z, W^+, W^+W^-$.

The dataset includes 302072 events, of which roughly 50\% is signal and 50\%
are background processes. All background processes have an equal number of events. 
There is no cut on the maximum number of objects and there is no order.    

The contents of the datasets (X\_train \& X\_val) are given below.
IMPORTANT: The specific line format of the dataset is as follows:

$$E_{T}^{miss}, \phi_{E_t}^{miss}, obj_1, E_1, p_{T1}, \eta_1, \phi_1, obj_2, E_2, p_{t2}, \eta_2, \phi_2, ...$$

Such that each object is represented by a string that starts with an identifier "$obj_n$", which is an
integer value representing a particular object in the event. The object identifier is
followed by its kinematic properties in the form of a four-vector containing the full 
energy "E" and the transverse momentum "p\_T" in units of MeV, as well as the pseudo-rapidity 
"$\eta$" and the azimuthal angle "$\phi$". The other three quantities are "weight" given by the cross-section
of the process divided by the total number of events generated. "$E_{T}^{miss}$" is the magnitude of the
missing transverse energy in units of MeV and "$\phi_{E_T}^{miss}$" is the azimuthal angle of the missing
transverse energy.

Since the length of the events is variable, the data is zero-padded to the largest number of objects
found in the events within the entire dataset. The dataset is fairly sparse and not pre-processed.

The relevant datasets are pytorch tensors with the following properties: \\

Name: X\_train, shape: [241657, 92], dtype: torch.float32,\\ 
Name: Y\_train, shape: [241657], dtype: torch.int64, \\
Name: X\_val, shape: [30272, 92], dtype: torch.float32, \\
Name: Y\_val, shape: [30272], dtype: torch.int64 \\

IMPORTANT: Each of these tensors are pre-loaded and available. \\

\begin{lstlisting}[language=Python]
    # ----------------  START HARNESS WRAPPER PREFIX (FOR CONTEXT)  ---------------- 
    # Environment: Python 3.12, PyTorch 2.6.0, Torch_Geometric 2.6.1, NumPy 2.2.3, SciPy v1.15.2, SciKit-Learn 1.6.1
    import os, sys, pickle, torch, torch_geometric, gc, json, importlib
    import pandas as pd
    import numpy as np
    import matplotlib.pyplot as plt
    from torch import nn
    from torch.utils.data import Dataset, DataLoader
    
    torch.manual_seed(42)                        
    os.environ["PYTHONHASHSEED"] = "42"
    SCRIPT_DIR = os.path.dirname(os.path.abspath(sys.argv[0]))
                            
    DATASET = {dataset_dict}
                           
    def load_data():
        X_train = pd.read_csv(DATASET["X_train"], dtype=np.float32).to_numpy(copy=False)
        Y_train = pd.read_csv(DATASET["Y_train"], dtype=np.int64).to_numpy(copy=False).ravel()
        X_val   = pd.read_csv(DATASET["X_val"], dtype=np.float32).to_numpy(copy=False)
        Y_val   = pd.read_csv(DATASET['Y_val'], dtype=np.int64).to_numpy(copy=False).ravel()
    
        gc.collect()
    
        return (torch.from_numpy(X_train), torch.from_numpy(Y_train),
                torch.from_numpy(X_val), torch.from_numpy(Y_val))
    
    class PairDataset(Dataset):
        def __init__(self, x, y):
            self.x = x
            self.y = y
    
        def __len__(self):
            return len(self.y)
            
        def __getitem__(self, idx):
        
            if isinstance(self.x, (tuple, list)) and all(torch.is_tensor(t) for t in self.x):
                return (tuple(t[idx] for t in self.x), self.y[idx])
            else:
                return (self.x[idx], self.y[idx])
    
    def _make_dataset(x, y):
        custom = globals().get("make_dataset", None)
        if callable(custom):
            ds = custom(x, y)
            if ds is not None:
                return ds
        return PairDataset(x, y)
    
    def make_loaders(X_train, Y_train, X_val, Y_val, *, batch=512, collate_fn=None, loader_cls=None):
        train_ds = _make_dataset(X_train, Y_train)
        val_ds   = _make_dataset(X_val , Y_val)
    
        if loader_cls is None: 
            loader_cls = DataLoader
    
        train_ld = loader_cls(train_ds, batch_size=batch, shuffle=True, num_workers=0, 
                            collate_fn=collate_fn)
        val_ld   = loader_cls(val_ds, batch_size=batch, shuffle=False, num_workers=0,
                            collate_fn=collate_fn)
    
        return train_ld, val_ld
    
    # ----------------  END HARNESS WRAPPER PREFIX (FOR CONTEXT)  ----------------                        
    # -------------------------- START OF LLM BLOCK ------------------------------
    
    ** Code Template **
    # <start code template>
    # 0. ---------- IMPORTS ----------
    # NOTE: Some imports (torch, nn, numpy, DataLoader) are already available (see prefix).
    # Only import extra std-lib modules, torch, scipy, sklearn (sub-)modules you actually use.
    # <LLM: Import modules>
    
    # 2. ---------- PRE-PROCESSING ----------
    class MyPreprocessor:
        #    Must implement:
        #   - fit(...)               -> self
        #   - transform(X: ???)      -> ???
    
        # DATA SPECIFICS
        # Total flat length per event (X_train & X_val): 92
        # Index  0 :  missing-ET magnitude  (E_T_miss)
        # Index  1 :  missing-ET azimuth    (phi_Et_miss)
        # Indices  2-6  : object 1  ->  obj_1, E_1, p_T1, eta_1, phi_1
        # Indices  7-11 : object 2  ->  obj_2, E_2 , p_T_2 , eta_2 , phi_2
        # ...
        # Indices 88-92 : object 18 ->  obj_18, E_18 , p_T_18 , eta_18 , phi_18
        # Global features       = 2
        # Per-object slice size = 5
        # Max objects encoded   = 18
    
        # TIPS
        # When modifying data features or feature engineering: annotate tensor size as comments after 
        # each tensor operation to reduce dimension mismatches.
    
        # REQUIREMENTS
        # IMPORTANT: All state must be picklable with the std-lib pickle module.
        # May allocate NumPy arrays or Torch tensors internally, but:
        # transform() must be deterministic.
        # Store only derived parameters needed for transform i.e. do not store the raw data
        # itself in the preprocessor object.
    
        # <LLM: Write code to preprocess the data> 
    
        def __init__(self):
            # <LLM: Define and initialize any stateful components here>
            pass
    
        def _raw_reshape(self, X):           
            # <LLM: Apply optional raw data reshaping logic here>
            return X # Returns identify by default
    
        # Uncomment to implement custom collate function.    
        # @staticmethod
        # def _collate_fn(batch: list):
        #    <LLM: Apply optional custom collate logic here>
        #    return None
    
        def make_loader_cfg(self):
            # Return dict or None.  If dict, evaluator uses it to rebuild loader:
            #{
            #   "loader_class": "torch.utils.data.DataLoader",
            #   "collate_fn": "self._collate_fn",
            #   "batch_size": 256,
            #   "shuffle": False,
            #   "num_workers": 0
            #}
            return None
    
        def fit(self, X, y=None):
            # <LLM: Extract statistics for fit transformers>
            return self
    
        def transform(self, X):
            # <LLM: Apply pre-processing logic>
            return X # must return an indexable, picklable object
    
        def fit_transform(self, X, y=None):
            self.fit(X, y)
            return self.transform(X)
    
    def make_preprocessor():
        return MyPreprocessor()
    
    # 2. ---------- MODEL DEFINITION ----------
    class BinaryClassifier(nn.Module):
        def __init__(self, sample_object):
            super().__init__()
            # <LLM: Define and initialize any stateful components here>
    
        # <LLM: optionally build extra layers here>
    
        def forward(self, *data):
            # <LLM: Define your model's forward pass here>
            pass
    
    def make_model(example_object):
        return BinaryClassifier(example_object)
    
    # 3. ---------- MODEL TRAINING ----------
    EPOCHS = 10   # <LLM: adjust if you wish>
    def train_model(model: nn.Module, train_loader, val_loader, epochs: int):
        # REQUIREMENTS 
        # Do NOT pass "verbose=" to any PyTorch scheduler (not supported in this image).
        # Must return trained_model, train_loss, val_loss, train_acc, val_acc
        # Implement early-stopping.
        # Forward signature must match.
    
        # <LLM: Write code to define training loop>
        # <LLM: Implement early stopping if possible>
        return trained_model, train_loss, val_loss, train_acc, val_acc
    
    # IMPORTANT: DO NOT execute the pipeline here - the harness will do that.
    # <end code template>
    
    
    # ---------------------------  END OF LLM-CODE BLOCK ---------------------------
    # ----------------  START HARNESS WRAPPER SUFFIX (FOR CONTEXT)  ---------------- 
    
    def _import_dotted(path: str):
        mod, name = path.rsplit(".", 1)
        module = importlib.import_module(mod)
        return getattr(module, name)
    
    def _plot(series_train, series_val, name, out_path):
        plt.figure()
        epochs = range(1, len(series_train) + 1)
        plt.plot(epochs, series_train, label=f"Train {name}")
        plt.plot(epochs, series_val,   label=f"Val {name}")
        plt.title(name); plt.xlabel("Epoch"); plt.legend()
        plt.savefig(out_path); plt.close()
    
    def _run(dryrun=False):
        # 1. Load & preprocess
        X_train, Y_train, X_val, Y_val = load_data()
        if dryrun:
            X_train, Y_train, X_val, Y_val = X_train[:200], Y_train[:200], X_val[:20], Y_val[:20]
        pre     = make_preprocessor().fit(X_train, Y_train)
        X_train = pre.transform(X_train)
        X_val   = pre.transform(X_val)
    
        collate = getattr(pre, "_collate_fn", None)
        cfg     = getattr(pre, "make_loader_cfg", lambda: None)() or {}
        loader_cls = _import_dotted(cfg["loader_class"]) if "loader_class" in cfg else None
        train_loader, val_loader = make_loaders(X_train, Y_train, X_val, Y_val, 
                                                batch      = cfg.get("batch_size", 512), 
                                                collate_fn = collate,
                                                loader_cls = loader_cls)
    
        # 2. Build model
        first_batch    = next(iter(train_loader))
        example_sample = first_batch[0]
        model          = make_model(example_sample)
    
        # 3. Train model
        n_epochs = 1 if dryrun else globals().get("EPOCHS", 10)
        try:
            trained_model, tr_loss, va_loss, tr_acc, va_acc = train_model(
                model, train_loader, val_loader, epochs=n_epochs)
        except Exception as e:
            print("ERROR during training:", e)
            raise
    
        # 4. Dry-run safety check
        if dryrun:
            sample, _ = first_batch
            try:
                _ = trained_model(*sample) if isinstance(sample, (tuple, list)) else trained_model(sample)
            except Exception as e:
                raise RuntimeError("Sanity-check forward pass failed") from e
            return
    
        # 5. Persist artefacts
        if not dryrun:
            base = os.path.splitext(os.path.basename(sys.argv[0]))[0].removeprefix("script_")
    
            pth_state   = os.path.join(SCRIPT_DIR, f"{base}_state.pt")
            pth_model   = os.path.join(SCRIPT_DIR, f"{base}_model.pkl")
            pth_preproc = os.path.join(SCRIPT_DIR, f"{base}_preproc.pkl")
    
            torch.save(trained_model.state_dict(), pth_state)
            with open(pth_model,   "wb") as f: pickle.dump(trained_model, f)
            with open(pth_preproc, "wb") as f: pickle.dump(pre,           f)
    
            # 6. Save plots
            _plot(tr_loss, va_loss, "Loss",     os.path.join(SCRIPT_DIR, f"{base}_loss.png"))
            _plot(tr_acc,  va_acc,  "Accuracy", os.path.join(SCRIPT_DIR, f"{base}_accuracy.png"))
    
        # 7. Write JSON Summary
        if not dryrun: 
            summary = {
                "epochs": n_epochs,
                "train_loss": tr_loss   if tr_loss else None,
                "val_loss":   va_loss   if va_loss else None,
                "train_acc":  tr_acc    if tr_acc else None,
                "val_acc":    va_acc    if va_acc else None,
            }
            print("#TRAIN_METRICS#" + json.dumps(summary))
    
    if "__main__" not in sys.modules:
        sys.modules["__main__"] = sys.modules[__name__]
    
    if __name__ == "__main__":
        _run(dryrun="--dryrun" in sys.argv)
    
    # ----------------  END HARNESS WRAPPER SUFFIX (FOR CONTEXT)  ---------------- 
\end{lstlisting}

 ** IMPORTANT: Your Challenge **
Write Python code for a binary classification model focusing on maximising the AUC using the code template above. 
You may freely choose any pre-processing methods and techniques as well as model architecture and training conventions.
Do absolutely everything in your power to achieve the highest possible AUC.

** Response Format **
Your response must strictly be python code. 
If you must wrap it, put it in a ```python fenced block and nothing else.
Your response must follow these rules:

1. Do not add any formatting, such as markdown, to the response. \\
2. Replace each "$\# <LLM: ...>$" comment, in the code template, with the required code. 
No placeholder should remain.\\
3. Before finalizing your answer, double-check that your code runs without errors and
meets all requirements (all functions implemented, correct tensor shapes, etc.).\\
4. To prevent dimensional mismatches make sure to annotate tensor sizes as comments.\\
5. IMPORTANT: Remember, your first, and most important priority is to produce 
(syntactically) correct code. Prioritise what you can implement reliably above all else. 
Then prioritise maximising the metric.

\subsubsection{Example Response: Google Gemini (23/06) AUC: 0.847}

\begin{lstlisting}[language=Python]
# 0. ---------- IMPORTS ----------
# NOTE: Some imports (torch, nn, numpy, DataLoader) are already available (see prefix).
# Only import extra std-lib modules, torch, scipy, sklearn (sub-)modules you actually use.
from sklearn.preprocessing import StandardScaler
from torch.optim.lr_scheduler import ReduceLROnPlateau
import copy


# 2. ---------- PRE-PROCESSING ----------
class MyPreprocessor:
    #    Must implement:
    #   - fit(...)               -> self
    #   - transform(X: ???)      -> ???

    # DATA SPECIFICS
    # Total flat length per event (X_train & X_val): 92
    # Index  0 :  missing-ET magnitude  (E_T_miss)
    # Index  1 :  missing-ET azimuth    (phi_Et_miss)
    # Indices  2-6  : object 1  ->  obj_1, E_1, p_T1, eta_1, phi_1
    # Indices  7-11 : object 2  ->  obj_2, E_2 , p_T_2 , eta_2 , phi_2
    # ...
    # Indices 88-92 : object 18 ->  obj_18, E_18 , p_T_18 , eta_18 , phi_18
    # Global features       = 2
    # Per-object slice size = 5
    # Max objects encoded   = 18

    # TIPS
    # When modifying data features or feature engineering: annotate tensor size as comments after 
    # each tensor operation to reduce dimension mismatches.

    # REQUIREMENTS
    # IMPORTANT: All state must be picklable with the std-lib pickle module.
    # May allocate NumPy arrays or Torch tensors internally, but:
    # transform() must be deterministic.
    # Store only derived parameters needed for transform i.e. do not store the raw data
    # itself in the preprocessor object.

    def __init__(self):
        # Define and initialize any stateful components here
        self.global_scaler = StandardScaler()
        self.object_scaler = StandardScaler()
        # Engineered features:
        # globals: log(1+E_T_miss), met_x, met_y
        self.n_global_feats = 3
        # objects: log(1+E), log(1+pT), eta, phi, px, py, pz
        self.n_obj_cont_feats = 7 

    def _reshape_and_feature_engineer(self, X: torch.Tensor):           
        # Apply optional raw data reshaping logic here

        # X: [N, 92]
        global_features_raw = X[:, :2]                       # [N, 2]
        objects_raw = X[:, 2:].reshape(-1, 18, 5)             # [N, 18, 5]

        # Create mask for valid objects (p_T > 0), pT is at index 2 of the 5-tuple
        mask = objects_raw[:, :, 2] > 1e-6 # Use a small epsilon for float comparison # [N, 18]

        # --- Feature engineering for global features ---
        E_T_miss = global_features_raw[:, 0]
        phi_E_T_miss = global_features_raw[:, 1]

        log_E_T_miss = torch.log1p(E_T_miss)
        met_x = E_T_miss * torch.cos(phi_E_T_miss)
        met_y = E_T_miss * torch.sin(phi_E_T_miss)

        global_features = torch.stack([log_E_T_miss, met_x, met_y], dim=1) # [N, 3]

        # --- Feature engineering for object features ---
        # The first feature is obj_id, which we treat as categorical
        obj_ids = objects_raw[:, :, 0]                       # [N, 18]
        E = objects_raw[:, :, 1]
        pT = objects_raw[:, :, 2]
        eta = objects_raw[:, :, 3]
        phi = objects_raw[:, :, 4]

        log_E = torch.log1p(E)
        log_pT = torch.log1p(pT)

        px = pT * torch.cos(phi)
        py = pT * torch.sin(phi)
        pz = pT * torch.sinh(eta)

        object_continuous_features = torch.stack([
            log_E, log_pT, eta, phi, px, py, pz
        ], dim=-1) # [N, 18, 7]

        # apply mask to zero out padded values
        object_continuous_features[~mask] = 0.0

        return global_features, obj_ids, object_continuous_features, mask


    def fit(self, X, y=None):
        # Extract statistics for fit transformers
        global_features, _, object_cont_features, mask = self._reshape_and_feature_engineer(X)

        # Fit global scaler
        self.global_scaler.fit(global_features.cpu().numpy())

        # Fit object scaler on valid objects only
        if mask.any():
            valid_object_features = object_cont_features[mask]
            self.object_scaler.fit(valid_object_features.cpu().numpy())

        return self

    def transform(self, X):
        # Apply pre-processing logic
        global_features, obj_ids, object_cont_features, mask = self._reshape_and_feature_engineer(X)

        # Scale global features
        global_features_np = global_features.cpu().numpy()
        scaled_global_features_np = self.global_scaler.transform(global_features_np)
        scaled_global_features = torch.from_numpy(scaled_global_features_np).to(global_features.dtype)

        # Scale object features
        scaled_object_features = torch.zeros_like(object_cont_features)
        if mask.any():
            valid_features = object_cont_features[mask]
            if valid_features.shape[0] > 0:
                scaled_valid_np = self.object_scaler.transform(valid_features.cpu().numpy())
                scaled_valid = torch.from_numpy(scaled_valid_np).to(valid_features.dtype)
                scaled_object_features[mask] = scaled_valid

        # Return a tuple of tensors to be handled by the default DataLoader
        return (
            scaled_global_features.float(),
            obj_ids.long(),
            scaled_object_features.float(),
            mask
        )

    def fit_transform(self, X, y=None):
        self.fit(X, y)
        return self.transform(X)

def make_preprocessor():
    return MyPreprocessor()

# 2. ---------- MODEL DEFINITION ----------
class BinaryClassifier(nn.Module):
    def __init__(self, sample_object):
        super().__init__()
        # Define and initialize any stateful components here
        globals, obj_ids, obj_feats, mask = sample_object

        n_global_features = globals.shape[1]
        n_obj_continuous_features = obj_feats.shape[2]

        # Infer max object ID from the first batch. This is a pragmatic approach.
        self.max_obj_id = int(obj_ids.max().item())

        d_model = 128
        n_head = 4
        n_layers = 4
        d_ff = 256
        dropout = 0.1
        self.obj_id_embedding_dim = 16

        self.obj_id_embedding = nn.Embedding(
            num_embeddings=self.max_obj_id + 2, # Add padding_idx and one for safety
            embedding_dim=self.obj_id_embedding_dim,
            padding_idx=0
        )

        self.continuous_input_proj = nn.Linear(
            n_obj_continuous_features, 
            d_model - self.obj_id_embedding_dim
        )

        encoder_layer = nn.TransformerEncoderLayer(
            d_model=d_model, nhead=n_head, dim_feedforward=d_ff,
            dropout=dropout, batch_first=True, activation=nn.GELU()
        )
        self.transformer_encoder = nn.TransformerEncoder(
            encoder_layer, num_layers=n_layers
        )

        self.cls_token = nn.Parameter(torch.randn(1, 1, d_model))

        self.classifier_head = nn.Sequential(
            nn.Linear(d_model + n_global_features, 256),
            nn.LayerNorm(256),
            nn.GELU(),
            nn.Dropout(0.2),
            nn.Linear(256, 128),
            nn.LayerNorm(128),
            nn.GELU(),
            nn.Dropout(0.2),
            nn.Linear(128, 1)
        )

    def forward(self, *data):
        # Define your model's forward pass here
        globals, obj_ids, obj_feats, mask = data

        B = globals.shape[0]  # Batch size

        # Project object features to d_model
        # obj_ids: [B, 18], obj_feats: [B, 18, 7]
        id_embeds = self.obj_id_embedding(obj_ids)                  # [B, 18, 16]
        cont_embeds = self.continuous_input_proj(obj_feats)        # [B, 18, 112]
        object_embeddings = torch.cat([id_embeds, cont_embeds], dim=-1) # [B, 18, 128]

        # Prepend CLS token for aggregation
        cls_tokens = self.cls_token.expand(B, -1, -1)               # [B, 1, 128]
        full_seq = torch.cat([cls_tokens, object_embeddings], dim=1) # [B, 19, 128]

        # Create transformer mask. Mask should be True for padded values.
        # Our input mask is True for valid values.
        cls_mask = torch.zeros(B, 1, dtype=torch.bool, device=mask.device) # CLS token is never padded
        transformer_mask = torch.cat([cls_mask, ~mask], dim=1) # [B, 19], True for padded element

        transformer_output = self.transformer_encoder(
            src=full_seq, 
            src_key_padding_mask=transformer_mask
        ) # [B, 19, 128]

        # Get CLS token output (the aggregated representation of all objects)
        cls_output = transformer_output[:, 0, :]                   # [B, 128]

        # Concatenate with global features
        # globals: [B, 3]
        combined_features = torch.cat([cls_output, globals], dim=1) # [B, 128 + 3]

        # Final classification head
        logits = self.classifier_head(combined_features).squeeze(-1) # [B]

        return logits

def make_model(example_object):
    return BinaryClassifier(example_object)

# 3. ---------- MODEL TRAINING ----------
EPOCHS = 30   # Adjust if you wish
def train_model(model: nn.Module, train_loader, val_loader, epochs: int):
    # REQUIREMENTS 
    # Do NOT pass "verbose=" to any PyTorch scheduler (not supported in this image).
    # Must return trained_model, train_loss, val_loss, train_acc, val_acc
    # Implement early-stopping.
    # Forward signature must match.

    device = torch.device("cuda" if torch.cuda.is_available() else "cpu")
    model.to(device)

    optimizer = torch.optim.AdamW(model.parameters(), lr=1e-3, weight_decay=1e-2)
    criterion = nn.BCEWithLogitsLoss()
    scheduler = ReduceLROnPlateau(optimizer, 'min', factor=0.1, patience=3)

    best_val_loss = float('inf')
    early_stopping_patience = 5
    patience_counter = 0
    best_model_state_dict = None

    train_loss_hist, val_loss_hist = [], []
    train_acc_hist, val_acc_hist = [], []

    for epoch in range(epochs):
        # --- Training Phase ---
        model.train()
        running_loss, correct_preds, total_preds = 0.0, 0, 0
        for data_tuple, labels in train_loader:
            inputs = [item.to(device) for item in data_tuple]
            labels = labels.to(device)

            optimizer.zero_grad(set_to_none=True)

            outputs = model(*inputs)
            loss = criterion(outputs, labels.float())

            loss.backward()
            optimizer.step()

            running_loss += loss.item() * labels.size(0)
            preds = torch.round(torch.sigmoid(outputs))
            correct_preds += (preds == labels).sum().item()
            total_preds += labels.size(0)

        epoch_train_loss = running_loss / total_preds
        epoch_train_acc = correct_preds / total_preds
        train_loss_hist.append(epoch_train_loss)
        train_acc_hist.append(epoch_train_acc)

        # --- Validation Phase ---
        model.eval()
        running_loss, correct_preds, total_preds = 0.0, 0, 0
        with torch.no_grad():
            for data_tuple, labels in val_loader:
                inputs = [item.to(device) for item in data_tuple]
                labels = labels.to(device)

                outputs = model(*inputs)
                loss = criterion(outputs, labels.float())

                running_loss += loss.item() * labels.size(0)
                preds = torch.round(torch.sigmoid(outputs))
                correct_preds += (preds == labels).sum().item()
                total_preds += labels.size(0)

        epoch_val_loss = running_loss / total_preds
        epoch_val_acc = correct_preds / total_preds
        val_loss_hist.append(epoch_val_loss)
        val_acc_hist.append(epoch_val_acc)

        scheduler.step(epoch_val_loss)

        # --- Early Stopping Logic ---
        if epoch_val_loss < best_val_loss:
            best_val_loss = epoch_val_loss
            patience_counter = 0
            best_model_state_dict = copy.deepcopy(model.state_dict())
        else:
            patience_counter += 1
            if patience_counter >= early_stopping_patience:
                break

    # Load best model weights before returning
    if best_model_state_dict:
        model.load_state_dict(best_model_state_dict)

    return model, train_loss_hist, val_loss_hist, train_acc_hist, val_acc_hist
\end{lstlisting}

\end{document}